\documentclass[journal]{IEEEtran}

\usepackage{lmodern}
\usepackage{cite}
\usepackage{float}
\usepackage{graphicx}
\usepackage{hyperref}
\usepackage[export]{adjustbox}
\usepackage[usenames]{color}
\usepackage{multirow}
\usepackage{bigstrut}
\usepackage{epsf}

\begin{document}

\newpage
\twocolumn
\pagenumbering{arabic}
 
\title{Individual performance calibration using physiological stress signals}

\author{Francisco~Hernando-Gallego~and~Antonio~Art\'es-Rodr\'iguez\\
	Department of Signal Theory and Communications, Universidad Carlos III de Madrid.\\
	Avenida de la Universidad 30, 28911 Legan\'es, Madrid, Spain.\\
	Email: fhernando@tsc.uc3m.es  antonio@tsc.uc3m.es}

\maketitle

\begin{abstract}	
	The relation between performance and stress is described by the Yerkes-Dodson Law but varies significantly between individuals. This paper describes a method for determining the individual optimal performance as a function of physiological signals. The method is based on attention and reasoning tests of increasing complexity under monitoring of three physiological signals: Galvanic Skin Response (GSR), Heart Rate (HR), and Electromyogram (EMG). Based on the test results with 15 different individuals, we first show that two of the signals, GSR and HR, have enough discriminative power to distinguish between relax and stress periods. We then show a positive correlation between the complexity level of the tests and the GSR and HR signals, and we finally determine the optimal performance point as the signal level just before a performance decrease. We also discuss the differences among signals depending on the type of test.
\end{abstract}

\section{Introduction}
\IEEEPARstart{D}{o} you remember the last time you felt genuinely nervous? Maybe you had a pressing deadline and very little time to write a report or perhaps you received an unpleasant e-mail you had to reply to. This tension can suppose the appearance of stress \cite{Hernandez2014}. In the literature, stress has been defined as a reaction from a calm state to an excited state in order to preserve the integrity of the organism \cite{Healey2000}. Hence, there are stressful situations where individuals must deal with changes in their condition (from a calm state to an excited state). This reaction is given by changes and pressures which provoke physical and physiological responses \cite{Selye1978}.

\begin{figure}
	\begin{center}
		\includegraphics[width=\linewidth]{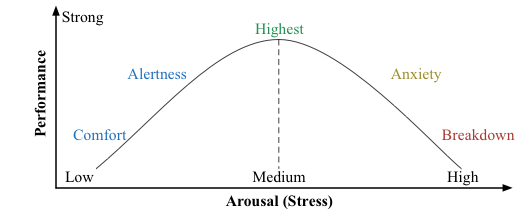}
		\caption{Yerkes-Dodson Law}
		\label{Yerkes-Dodson}
	\end{center}
\end{figure}

The presence of stress has been associated with a decrease in performance \cite{Matthews2000}. According to the Yerkes-Dodson Law \cite{Yerkes1908}, This achievement is a function of arousal, as shown in Fig. \ref{Yerkes-Dodson}. Performance increases with arousal when any individual feel relaxed, then reaches its peak at the highest arousal level because the patient be involved in the task, and decreases when the individual feel in a breakdown or anxiety situation. 

The aim of this article is to use the Yerkes-Dodson Law to design an individual performance curve using physiological responses. We need to explore all the possible stress values (from a comfort to a breakdown state) to obtain the performance of a user. We already know that humans reacts to stress situations, next questions we need to answer are when these situations happen and how we can measure this stress.

Stress reactions are different for each person, but present similar behaviours doing the same exercise. Different tests have been used in the literature for stress detection. P. Renaud and J. Blondin \cite{Renaud1997} used Stroop color test \cite{jensen1966stroop}, a recognition test with in congruent questions to obtaining the individual performance based on number of correct answers. E. Jovanov et al. \cite{Jovanov2003} used specific military personal training to know the performance of a soldier based on time reaction.

Arousal is also closely related to subject's performance in mental tasks. There are different ways to determine the level of arousal that an individual is achieving while performing activities. Two subjective approaches: SAGAT and NASA-TLX, have been used in aeronautic operations. Using the first one, the fly controller was then required to answer a written questionnaire in order to compare his mental model with reality. Questions were selected according to the spatial knowledge. On the other hand, NASA Task Load Index (NASA-TLX) is a multi-dimensional scale designed to estimate the controller's workload while performing a task. It consists of six subscales that represent the subjective workload experienced by the controller. These scales involve mental, physical and temporal demand, effort, frustration and performance.

Emotional states provoke changes in different physiological signals that can be measured in order to obtain information about the mental state of the individual. Various methods for detecting stress levels by monitoring physiological signals in different status are already presented in the literature. D. Wu et al. \cite{Wu2010} captured in real-time physiological responses: Galvanic Skin Response (GSR), Respiration, ECG and Electroencephalogram (EEG). They captured these signals for identification and classification into an optimal arousal as indicated the Yerkes-Dodson Law taking into account the user performance. Another case was Zhai and Barreto \cite{Zhai2005}, who's acquired different affective features: GSR, Blood Volume Pressure (BVP), Pupil Diameter (PD) and Skin Diameter (ST) to differentiate states (relax or mental stress) in computers user.

The main scope of this work is to analyse physiological signals and determinate when an individual starts to suffer a decrease in performance, according to the Yerkes-Dodson curve. By eliciting each subject from a low level arousal to a breakdown situation, it is possible to find a proper calibration for the Yerkes-Dodson Law stress detection.

\section{Methodology}
Our proposal consists of eliciting each subject to simulate the arousal of Yerkes-Dodson Law. To explore all the stress possibilities, our method is based two different test: one of visual attention and another of logical reasoning. Each individual starts in a comfort situation, continuing with an alertness and optimal performance state, and finishing with an anxiety and breakdown condition. If we go across the  obtainedresults, its calibrates each individual performance because we have all the possible arousal points. Therefore we only have to make a relation with the achievement performance.

We propose to detect stress signs in real time and acquisition in a non-intrusive way with wearable devices to minimize the inconvenience caused by the monitoring. Taking into account different biomedical responses, our proposal is to capture signals using  wireless sensors which acquire signal based in movement, heart beats and Electrodermal activity (EDA).

\subsection{Data Acquisition - Hardware Equipment}
Three SHIMMER (Shimmer Research, Dublin, Ireland) sensors were used for data acquisition.

\begin{figure}
	\begin{center}
	\includegraphics[width=.8\linewidth, frame]{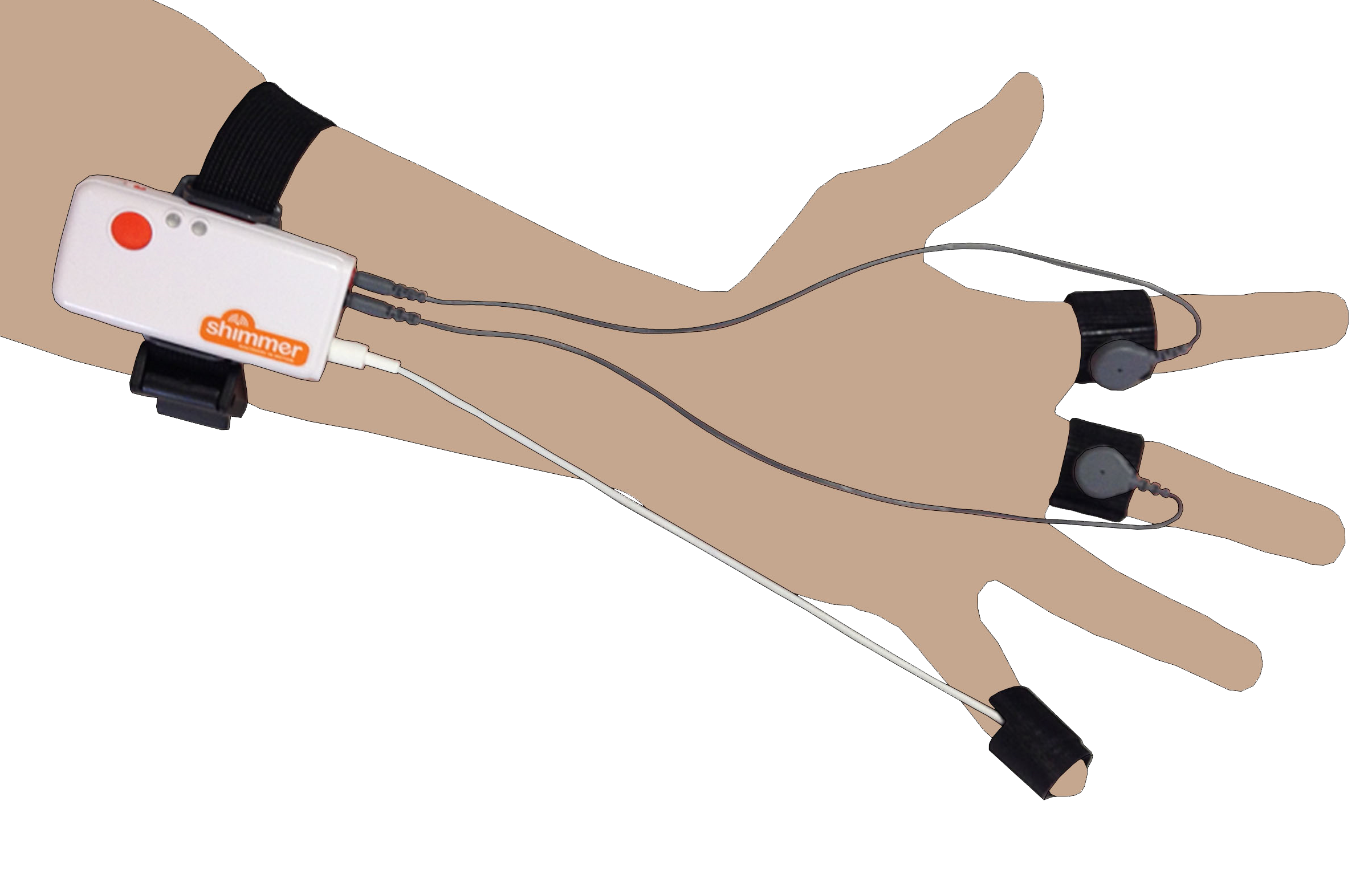}
		\caption{Positioning used for GSR and BVP signals acquiring electrodes and optical pulse sensing probe}
		\label{shimmerBVP}
	\end{center}
\end{figure}

A Shimmer3 `GSR Unit' sensor was used to capture BVP and GSR signals. On the left hand, the BVP optical pulse sensor was placed on the palmar surface of the pinky finger \cite{Shimmer2014ppg}. For GSR signals two electrodes on the palmar surface of the middle and index fingers were placed \cite{Shimmer2014gsr} as shown in Fig. \ref{shimmerBVP}.

\begin{figure}
	\begin{center}
		\includegraphics[width=.8\linewidth, frame]{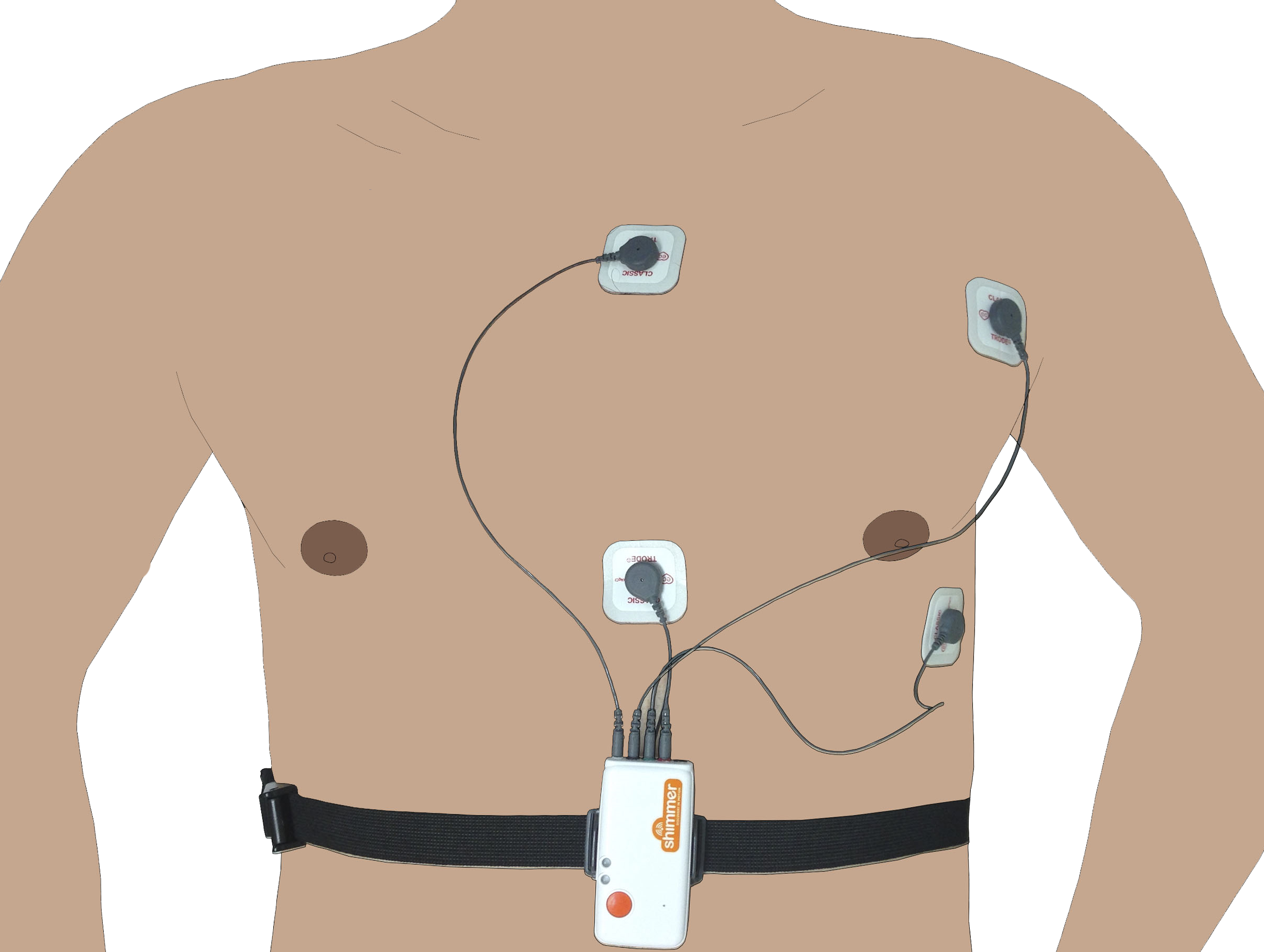}
		\caption{Positioning used for the ECG measurement electrodes}
		\label{shimmerECG}
	\end{center}
\end{figure}

Another sensor model, Shimmer3 `ExG Unit', was used to capture EMG and ECG signals. In the case of ECG, Fig. \ref{shimmerECG} shows how the electrodes were positioned on the chest \cite{Shimmer2014ecg}. 

\begin{figure}
	\begin{center}
		\includegraphics[width=.8\linewidth, frame]{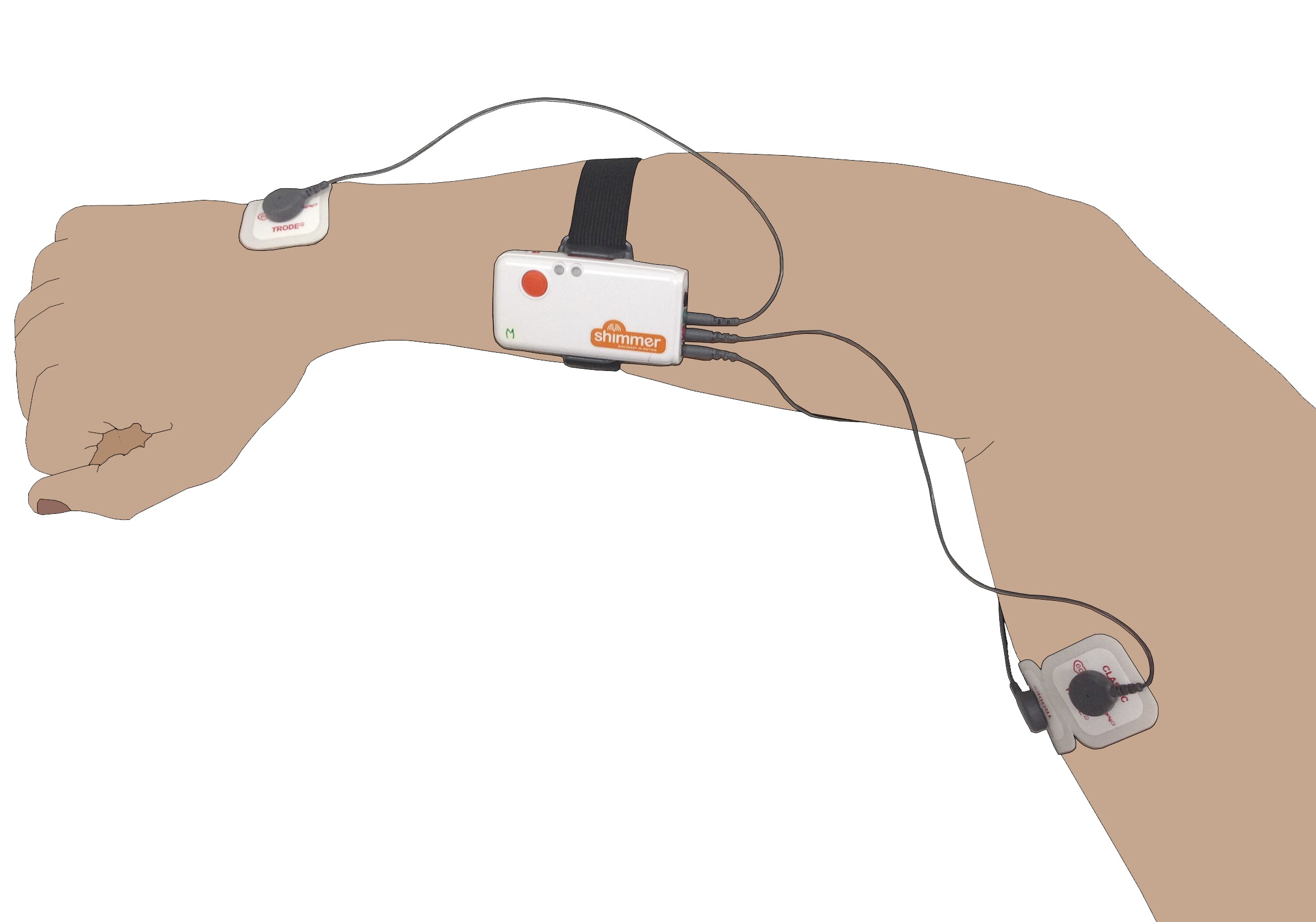}
		\caption{Positioning used for the EMG measurement electrodes}
		\label{shimmerEMG}
	\end{center}
\end{figure}

In order to capture EMG signals, Fig. \ref{shimmerEMG} shows an example of right arm electrode layout. Two electrodes were placed in parallel with the muscle fibres of the biceps, near the centre of the muscle and the reference electrode  an electrically neutral point of the body, as far away as reasonably possible from the muscle being measured \cite{Shimmer2014emg}.

\subsection{Experimental Protocol}
An experimental test was developed to capture physiological signals. To asses the effectiveness, we collected experimental data from two states conditions: mental stress and relaxation.

Two different tests have been designed to elicit stress: stroop color word test \cite{jensen1966stroop} and mathematical operations, whereas the relaxation condition consisted on deep breathing exercises. With Stroop test we found the reaction activity of each individual, while with the maths operations try to evaluate the capacity of concentration. Physiological measures of GSR, ECG, EMG and BVP were continuously recorded throughout exposure to the environment. In this study, we propose the protocol shown in Table \ref{Tab_Overview}:

\begin{table}[H]
	\begin{center}
		\begin{tabular}{c|c| c}
			Scenario & Test & Duration (min) \\
			\hline
			I  & breathe deeply & 4   \bigstrut \\
			II &  color naming  & 4   \bigstrut \\
			III  & breathe deeply & 4   \bigstrut \\
			IV  & maths operations & 5   \bigstrut \\
			V  & breathe deeply & 3   \bigstrut \\
		\end{tabular}
		\vspace*{.3cm}\caption{Experimental Protocol}\label{Tab_Overview}
	\end{center}
\end{table}

The color naming test consisted of 105 slides. Seven colors (yellow, red, green, blue, black, white and orange) were displayed with equal probability. The subjects had to say each color that appeared in the display under time pressure.

The maths operations test consisted of 49 slides. Four operations (addition, substation, multiplication and division) were used with equal probability. The subjects had to write the answer in a paper under time pressure.

For both test, the answers are divided in seven parts, increasing from low to high level of difficulty, trying to approach the Yerkes-Dodson arousal.

Finally, subjects were asked to breathe deeply at a pace of 0.1 Hz (breathe in for 4 seconds, breathe out for 6 seconds). 

A total of 15 subjects participated in this experiment. Strict exclusion criteria were enforced so as to minimize the possible confounding effects of additional factors known to adversely impact a person's ability to process information.  We interleaved deep breathing between stress test to allow subjects to recover between consecutive stressors.

\subsection{Signal processing}

Captured signals were processed to obtain a relation between stress and personal performance. A briefly description for each one is described below.

\subsubsection{ECG} information about heart rate can be extracted subtracting the times which two consecutively maximums happen. A real-time peak ECG detection method was applied, based on comparison between absolute values of sum differentiated peaks and a simple threshold detection \cite{Pan1985}.

\subsubsection{BVP} this signal offers information about the heart beats and about the relative constrictions of the blood vessels. In order to measure the heart rate, we only had to calculate the distance between each maximum in seconds. Following the ECG method, we developed a peak BVP detector with similar threshold, conform to BVP signals.

\subsubsection{EMG} when the body suffers some stimulus, muscle activation reacts, which supposes an increment in the current measured through muscles. Therefore, by measuring muscle activity, these current increments corresponds to stress situations that will be able to be detected. Fast movements of the biceps were detected and discarded electrodes fluctuations and possible noise.

\subsubsection{GSR} skin conductance responses can be divided two components: tonic component and phasic component.
\begin{itemize}
	\item Tonic Component: it is the part of the skin conductance signal that changes very slowly and without any responses. It is also called Skin Conductance Level (SCL).
	\item Phasic Component: it is the part of the signal that is produced by a stimulus. It is also called Skin Conductance Response (SCR).
\end{itemize}
SCR is an important measure due to its simple waveform, its ability to indicate a response to single stimuli. Considering previous models \cite{Lim1997}, \cite{Benedek2000} the signal was processed  by an autoregressive deconvolution method to get the response of the sweat glands of skin by sudomotor nerve activity in SCR signals.

\section{Results}

\subsection{Feature extraction}
\begin{figure}[H]
	\begin{center}
		\includegraphics[scale=.4]{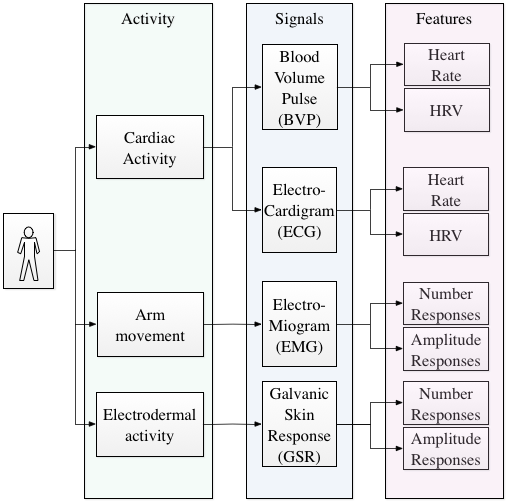}
		\caption{Physiological features extracted} \label{Features}
	\end{center}
\end{figure}

Different levels of arousal was implemented to determine the performance level used. Next question is whether or not these arousal levels could be recognized accurately from physiological signals.
As it was mentioned before, these signals are used in order to extract information about the mental state of an individual. Eight features were extracted from the data, including two from each signal, as shown in Fig. \ref{Features}. 

From BVP and ECG signals, statistical HR and HR variability were calculated, while the number of response and their amplitude of the EMG and GSR signals were captured.

\subsection{Stress signs in different scenarios}

First of all, compare the results obtained in the two scenarios: mental stress and relaxation was analysed and the following considerations were extracted:

\begin{itemize}
	\item Features obtained for the ECG and BVP were the same and the results obtained of HR and HRV  were similar. The ECG feature detector had more precision so the cardiac activity extracted from BVP signals were discarded leaving the ECG as the only source of heart activity to consider in the rest of the paper.
	
	\item Analysing the EMG results, no coherence with the scenarios set up could be extracted. Due to the experimental protocol followed to generate stress in the subjects, no abrupt movements were expected. This indicates that information obtained from the EMG has no significant diagnostic value when compared to the other variables.
\end{itemize}

\begin{figure}
	\begin{center}
		\includegraphics[width=\linewidth]{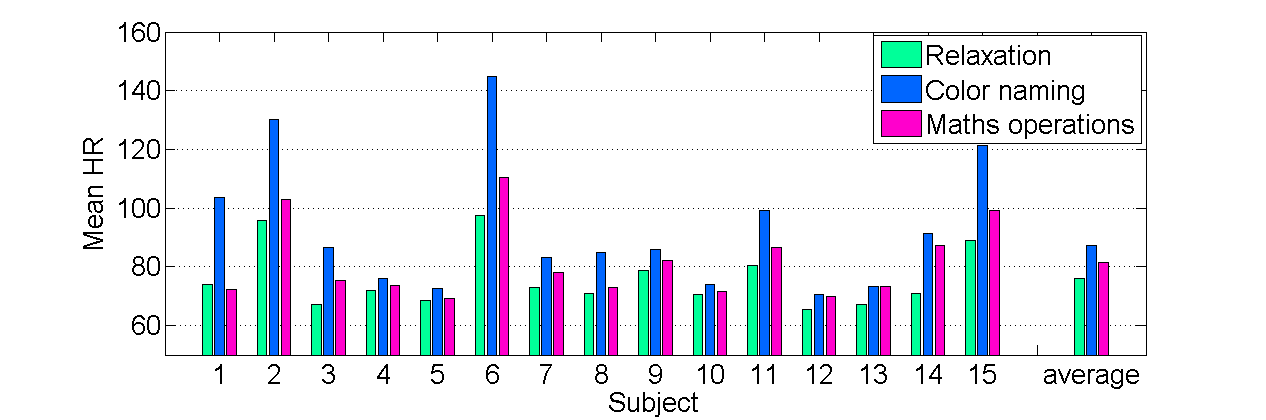}
		\caption{Average HR for each individual and global HR average}
		\label{fig_4_ecgMean}
	\end{center}
\end{figure}

\begin{figure}
	\begin{center}
		\includegraphics[width=\linewidth]{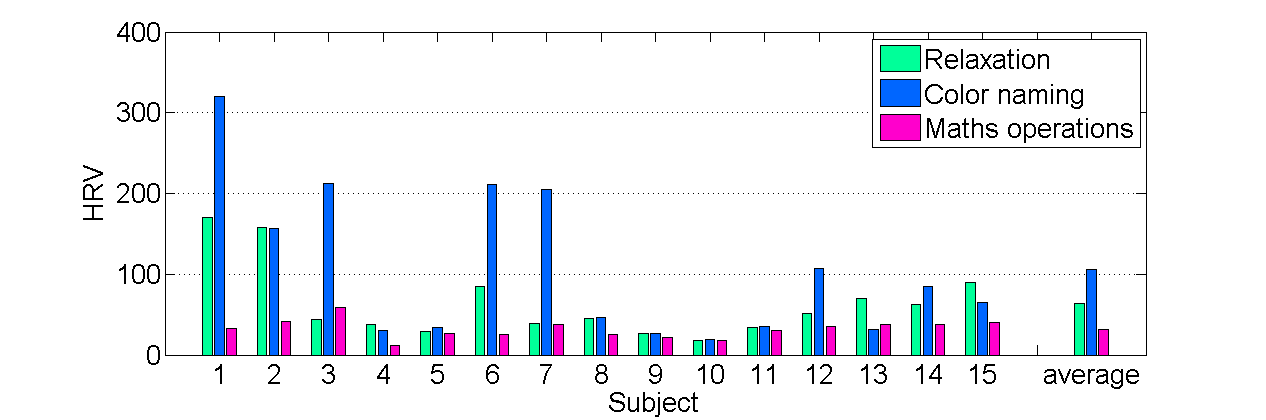}
		\caption{Average HR Variability for each individual and global HRV average}
		\label{fig_4_ecgStd}
	\end{center}
\end{figure}

\begin{figure}
	\begin{center}
		\includegraphics[width=\linewidth]{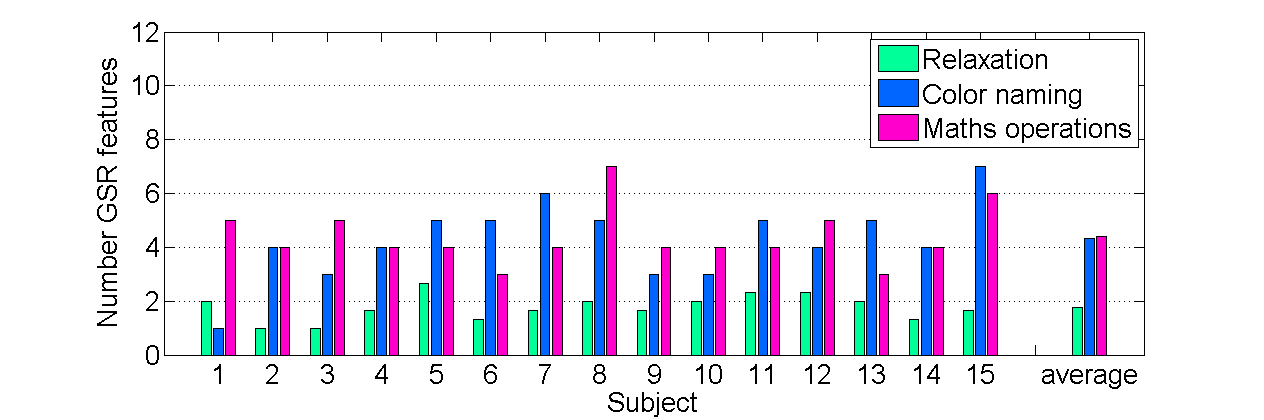}
		\caption{Number of GSR peaks for each individual and global GSR peaks average}
		\label{fig_4_gsrNumber}
	\end{center}
\end{figure}

\begin{figure}
	\begin{center}
		\includegraphics[width=\linewidth]{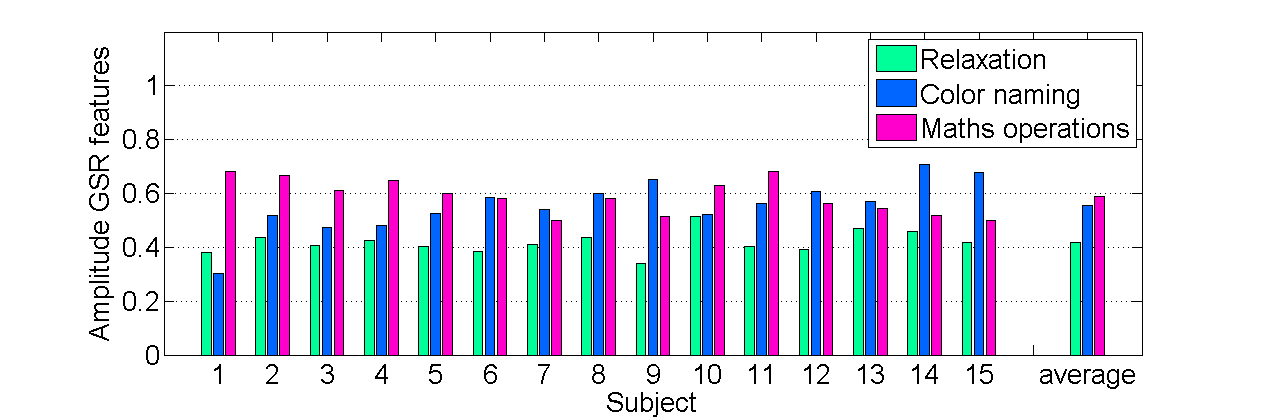}
		\caption{Amplitude of GSR peaks for each individual and global GSR peaks average in two different states}
		\label{fig_4_gsrAmplitude}
	\end{center}
\end{figure}

\begin{figure}
	\begin{center}
		\includegraphics[width=\linewidth]{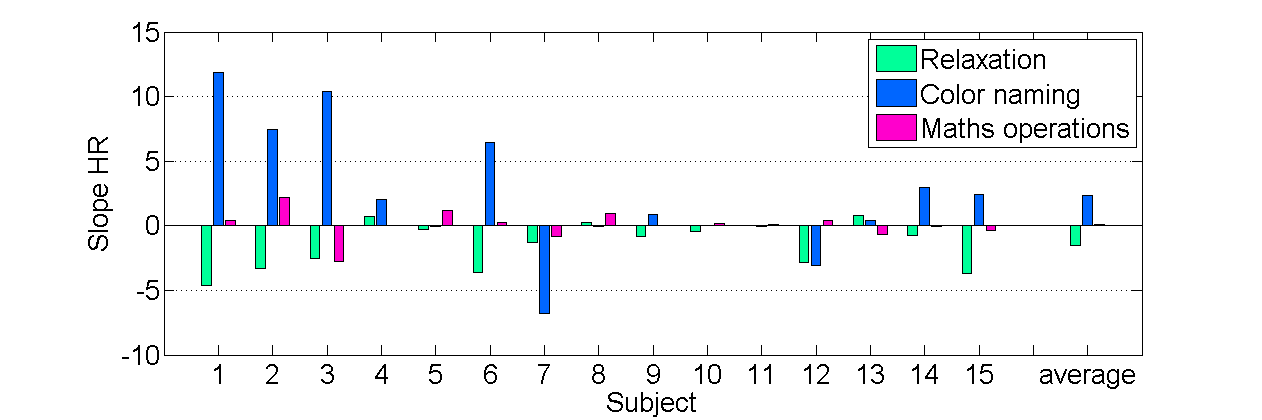}
		\caption{HR slope for each individual and global HR slope average}
		\label{fig_4_slopeECG}
	\end{center}
\end{figure}

\begin{figure}
	\begin{center}
		\includegraphics[width=\linewidth]{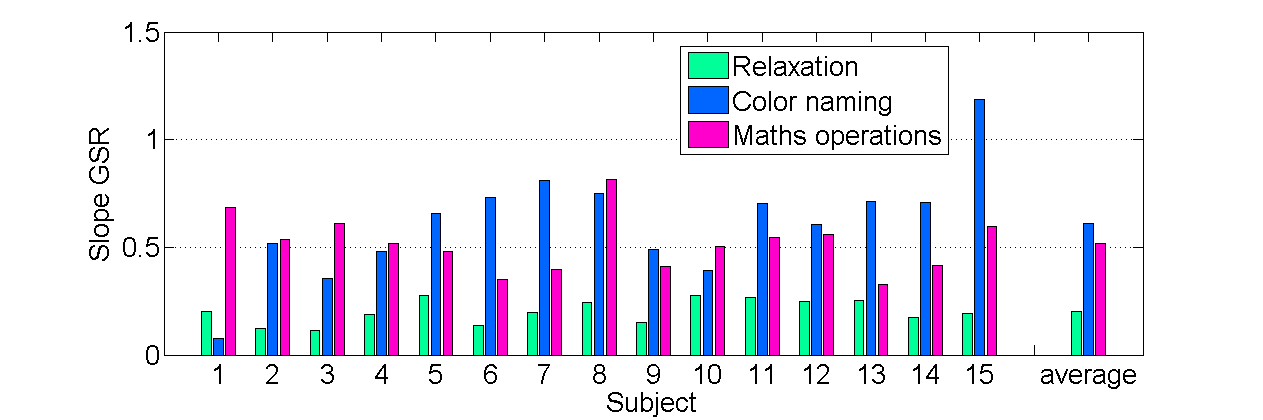}
		\caption{Number of peaks slope for each individual and global HR slope average}
		\label{fig_4_slopeGSR}
	\end{center}
\end{figure}

Taking into account these considerations, the results of the ECG and GSR features are shown for each individual. Fig. \ref{fig_4_ecgMean} and Fig. \ref{fig_4_ecgStd} represents the HR average and HR standard deviation for each scenario, besides Fig. \ref{fig_4_gsrNumber} and Fig. \ref{fig_4_gsrAmplitude} shows the number of GSR signs found in the experiment and their average amplitude. Moreover, Fig. \ref{fig_4_slopeECG} presents the personal slope that fits the best HR in a least-square sense of degree one and Fig. \ref{fig_4_slopeGSR} shows as well the cumulative sum of the GSR amplitude peaks. The slope value means the increment of the HR in the case of ECG, so a great slope indicates that the increment is high. Otherwise, it seems that the increment was low or even a decrease. For GSR peaks, a high slope level indicates that the subject suffered more stress in anxiety and breakdown levels than in other arousal level.

\subsection{Performance identification}
\begin{figure}
	\begin{center}
		\includegraphics[width=\linewidth]{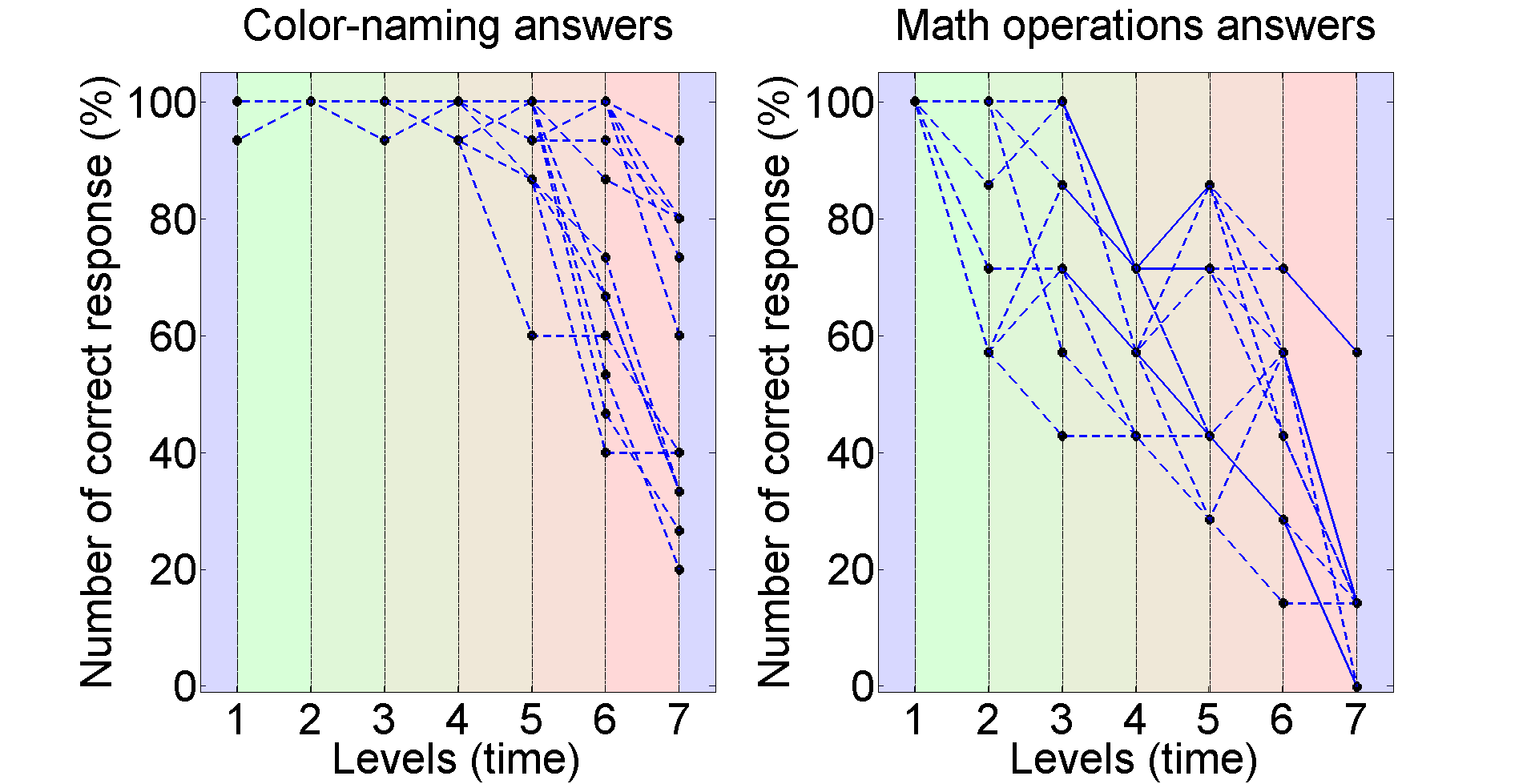}
		\caption{Number of correct responses for the 15 subjects. a) Scenario II: color naming test. b) Scenario IV: Maths operations}
		\label{Results}
	\end{center}
\end{figure}

Each subjects's performance was evaluated using the number of correct responses, which is the number of times that the subject indicate the correct color or answered a maths operation correctly under time pressure. As is given in \cite{Yerkes1908}, the number of correct answers showed good correlation with the arousal level that we want to obtain.  We partitioned these two test in seven groups, where the first group was the easier (comfort) and the seventh was the most difficult (breakdown). The individual results are shown in Fig. \ref{Results}, where the percentage of correct response were represented for each level.

In summary, when the difficult of the scenario increases, the number of correct response, which have been used as a performance measure, decreases, as was expected in theory.

The mean and standard deviation of the number of correct responses for all 15 subjects are shown in Table \ref{Tab:ColorNaming} and Table \ref{Tab:Mathoperat} to present a statistical reference for the Fig. \ref{Results}.

\renewcommand{\multirowsetup}{\centering} 
\setlength{\tabcolsep}{10pt}
\begin{table}[H]
	\begin{center}
		\begin{tabular}{|c|@{} c @{}|@{} c @{}|@{} c @{}|@{} c @{}|@{} c @{}|@{} c @{}|@{} c @{}|}
			\hline
			& \multicolumn{7}{p{5cm}|}{\centering Levels} \bigstrut \\
			\cline{2-8} & \multicolumn{1}{c|}{1} & \multicolumn{1}{c|}{2} & \multicolumn{1}{c|}{3} & \multicolumn{1}{c|}{4} & \multicolumn{1}{c|}{5} & \multicolumn{1}{c|}{6} & \multicolumn{1}{c|}{7} \bigstrut \\ \hline
			Mean (\%) & 99.55 & 100 & 99.55 & 98.22 & 92.88 & 76.88 &  52.44  \bigstrut \\
			Stad (\%) & 1.72 & 0 & 1.72 & 3.05 & 10.82 &  21.50 & 23.88 \bigstrut \\
			\hline
		\end{tabular}
	\end{center}
	\caption{Color naming test}\label{Tab:ColorNaming}
\end{table}

\renewcommand{\multirowsetup}{\centering} 
\setlength{\tabcolsep}{10pt}
\begin{table}[H]
	\begin{center}
		\begin{tabular}{|c|@{} c @{}|@{} c @{}|@{} c @{}|@{} c @{}|@{} c @{}|@{} c @{}|@{} c @{}|}
			\hline
			& \multicolumn{7}{p{5cm}|}{\centering Levels} \bigstrut \\
			\cline{2-8} & \multicolumn{1}{c|}{1} & \multicolumn{1}{c|}{2} & \multicolumn{1}{c|}{3} & \multicolumn{1}{c|}{4} & \multicolumn{1}{c|}{5} & \multicolumn{1}{c|}{6} & \multicolumn{1}{c|}{7} \bigstrut \\ \hline
			Mean (\%) & 100 & 80.61 & 82.65  & 60.20 &  56.12 & 43.87 &  15.30  \bigstrut \\
			Stad (\%) & 0 & 19.89 & 18.73 &  11.45 & 21.31 &  18.12 &  18.97 \bigstrut \\
			\hline
		\end{tabular}
	\end{center}
		\caption{Mathematical operations test}\label{Tab:Mathoperat}
\end{table}

\subsection{Personal performance}
Finally, to find a relation with the Yerkes-Dodson Law, each subject was studied separately to determinate the point where its performance start decreases. The results shown in Table \ref{Tab:Features} presents the stress features recognition when each individual decrease his performance in both test (color naming test is called Test1 and maths operations is named Test2). Three values are analysed:

\begin{enumerate}
	\item Both test were divided in seven levels, and this analysis shows in which level each individual start to decrease the number of correct responses.
	\item Taking into account this point of decrease, the increment of HR between the start of the test and this point was calculated.
	\item As the same way, the number of GSR peaks response were summed until the performance decreases.
\end{enumerate}

These results were collected in a table because the two stress features of stress had different processing. HR is a continuous function that varies in seconds, but in the other hand, the duration of the GSR peaks is approximately 15 seconds, so the count is discrete.

\renewcommand{\multirowsetup}{\centering} 
\setlength{\tabcolsep}{10pt}

\begin{table}
	\begin{center}		
		\begin{tabular}{|p{.7cm}||p{.4cm}|p{.4cm}|p{.4cm}|p{.4cm}|p{.4cm}|p{.4cm}|p{.4cm}|}
			\hline
			Subject & \multicolumn{2}{c|}{\centering Level decr.} & \multicolumn{2}{c|}{\centering HR incr.} & \multicolumn{2}{c|}{\centering GSR peaks} \bigstrut \\
			\cline{2-7} & Test1 & Test2 & Test1 & Test2 & Test1 & Test2 \bigstrut \\
			\hline
			1   & 6 & 6 & 70.45 &  6.41 & 2 & 4 \bigstrut \\
			2   & 5 & 5 & 35.21 & 17.89 & 3 & 3 \bigstrut \\
			3   & 5 & 4 & 44.31 &  3.41 & 3 & 3 \bigstrut \\
			4   & 6 & 5 & 18.79 &  2.21 & 3 & 2 \bigstrut \\
			5   & 6 & 6 &  1.21 &  5.12 & 4 & 3 \bigstrut \\
			6   & 6 & 5 & 49.84 &  9.12 & 4 & 2 \bigstrut \\
			7   & 7 & 4 &  6.28 & 23.15 & 6 & 1 \bigstrut \\
			8   & 6 & 5 & 24.15 &  8.09 & 4 & 6 \bigstrut \\
			9   & 6 & 5 & 14.99 & 19.01 & 3 & 4 \bigstrut \\
			10  & 7 & 6 & 11.42 & 16.20 & 3 & 4 \bigstrut \\
			11  & 6 & 4 & 25.15 & 11.15 & 4 & 3 \bigstrut \\
			12  & 6 & 4 & 12.93 & 17.17 & 3 & 3 \bigstrut \\
			13  & 5 & 5 & 14.19 & 20.15 & 3 & 2 \bigstrut \\
			14  & 7 & 6 & 39.70 & 26.15 & 4 & 3 \bigstrut \\
			15  & 6 & 5 & 44.91 & 37.31 & 6 & 4 \bigstrut \\
			\hline
		\end{tabular}
	\end{center}
	\caption{Individual features when performance decreased}\label{Tab:Features}
\end{table}

\section{Discussion and Conclusion}
An individual performance calibration have been designed. We used two tests for assessment of affective functioning to identify the performance level in mental tasks. The experiment consist of three wireless sensors that acquire signals: ECG, EMG, BVP and GSR. The system was validated through a series of test that the subject was elicited to stress or relax situations. The number of correct responses were used to distinguish among different arousal levels. 

BVP signals where discarded because only the information extracted from them is similar to the obtained from the ECG measurements. In the other hand, movement activity were discarded because after the experiment we saw that the results were not relevant stress information in this specific experiment. The results form the research outlined shows a promising correlation between the emotional stress and the ECG and GSR signals monitored as it is shown in Fig. \ref{fig_4_ecgMean}, Fig. \ref{fig_4_ecgStd}, Fig. \ref{fig_4_gsrNumber} and Fig. \ref{fig_4_gsrAmplitude}. Our results show that, under controlled conditions the ECG and GSR features can differentiate between relaxed and stressed user states, as elicited by incongruent Stroop simulation and maths operation test. We can conclude:

\begin{itemize}
		\item Two of the signals, GSR and ECG, have discriminative power to distinguish between relaxation and stress periods
		\item There exists a positive correlation between the complexity level of the tests and the GSR and ECG signals
		\item The part just before the number of correct responses decrease is considered the highest performance level of the Yerkes-Dodson Law
\end{itemize}

Besides, Fig. \ref{fig_4_slopeECG} and Fig. \ref{fig_4_slopeGSR} has performed to have a relation between the curve of Yerkes-Dodson Law and our proposal experiment. This two figures shows that the slope is greater in the metal stress than relaxation situations so it seems that each patient has more stress reactions in a difficult level than in a easy level.

Allow for Fig. \ref{Results}, Table \ref{Tab:ColorNaming} and Table \ref{Tab:Mathoperat}, we can conclude that the subjects performance starts to decrease sooner in the color naming than in the maths operation test.

HR has more increment more in reaction naming tests, otherwise the reasoning test need more concentration in average for GSR peaks at it is shown in Table \ref{Tab:Features}.

Additionally, Table \ref{Tab:Features}, can be used to compare the results obtained from the individuals performance identification. We can observe how much increase in HR and GSR peaks each individual showed before the performance decrease. Taking into account that one of the main aim of this paper is to study individuals response to stress situations, the mentioned table clearly shows how the subjects psychological reactions behave in different ways.

An emphasis of the system presented in this paper was the use of signals that can be collected under normal conditions of computer usage. All the sensor used are non-intrusive. We expect that, eventually, all sensors could be build into standard pieces of a computer, such as a smart-watch with a BVP and GSR acquisition system. The experiments cannot be utilised for the same subj ects because they learn the answers and the results were unexpected.

This research confirms the potential of integrated digital signal processing  to differentiate key affective states of any user from a suitable set of physiological responses.


\bibliographystyle{ieeetr}

\end{document}